\documentclass{ifacconf}

\usepackage{graphicx}      
\usepackage{natbib}        

\usepackage[T1]{fontenc}
\usepackage{xcolor}
\usepackage{amsmath,amssymb,amsfonts}

\usepackage{nameref}
\usepackage{bbm}
\usepackage{algorithmic}
\usepackage{textcomp}
\usepackage{dsfont}
\usepackage{float}
\usepackage{todonotes}
\usepackage{subcaption}

\usepackage{enumitem}

\usepackage{tikz} 
\usepackage{tikz-network}
\usetikzlibrary{arrows.meta,calc, angles, quotes}
\usepackage{subcaption}
\usepackage{standalone}
\usetikzlibrary{arrows,intersections,calc,positioning, angles, quotes}
\usetikzlibrary{decorations.markings}
\usetikzlibrary{shapes.misc}
\usetikzlibrary{patterns}
\usepackage{amsfonts}
\usepackage{mathtools}
\usepackage{upgreek}
\usepackage{float}
\usepackage{empheq}
\usepackage[theorems,skins]{tcolorbox}
\newtcolorbox{mymathbox}[1][]{colback=white, sharp corners, #1}
\usepackage[dvipsnames]{xcolor}
\usepackage{mathrsfs}

\let\theoremstyle\relax

\usepackage{amsthm}

\theoremstyle{definition}   
\newtheorem{definition}{Definition}

\theoremstyle{plain}
\newtheorem{proposition}{Proposition}

\newtheorem{theorem}{Theorem}
\newtheorem{lemma}{Lemma}
\newtheorem{example}{Example}

\newcommand{\Null}{\mathrm{Null}}
\newcommand{\IM}{\mathrm{IM}}
\newcommand{\diag}{\mathrm{diag}}
\newcommand{\Aut}{\mathrm{Aut}}

\global\long\def\G{\mathcal{G}}
\global\long\def\E{\mathcal{E}}
\global\long\def\V{\mathcal{V}}
\global\long\def\qq{\mathcal{Q}}

\global\long\def\C{\mathcal{C}}
\global\long\def\rr{\mathcal{R}}
\global\long\def\R{\mathbb{R}}
\global\long\def\ar{\mathscr{R}}
\global\long\def\fr{\mathscr{F}}
\global\long\def\sr{\mathscr{S}}

\global\long\def\l{\ell}

\begin{document}
\begin{frontmatter}

\title{Symmetry-Based Formation Control on Cycle Graphs Using Dihedral Point Groups\thanksref{footnoteinfo}} 

\thanks[footnoteinfo]{This work was supported by the Israel Science Foundation grant no. 453/24 and the Gordon Center for Systems Engineering.}

\author[First]{Zamir Martinez} 
\author[First]{Daniel Zelazo} 

\address[First]{Stephen B. Klein Faculty of Aerospace Engineering, Technion - Israel
Institute of Technology, Haifa, Israel  (e-mail:
z.m@campus.technion.ac.il, dzelazo@technion.ac.il).}

\begin{abstract}                
This work develops a symmetry-based framework for formation control on cycle graphs using Dihedral point-group constraints. We show that enforcing inter-agent reflection symmetries, together with anchoring a single designated agent to its prescribed mirror axis, is sufficient to realize every $\C_{nv}$-symmetric configuration using only $n-1$ communication links. The resulting control laws have a matrix-weighted Laplacian structure and guarantee exponential convergence to the desired symmetric configuration. Furthermore, we extend the method to enable coordinated maneuvers along a time-varying reference trajectory. Simulation results are provided to support the theoretical analysis.
\end{abstract}


\end{frontmatter}

\section{Introduction}
The study of distributed formation control in multi-agent systems (MAS) has drawn increasing attention in recent years. It plays a central role in many application domains, including distributed sensing architectures (\cite{Chenyu2024}), motion coordination for autonomous vehicle platoons (\cite{Dai2018,Chung2018}), and swarm coordination for mapping, surveillance, and localization (\cite{Tron2016}). A fundamental requirement in these applications is to steer the agents toward a prescribed formation using only local interactions. This is typically accomplished by exploiting rigidity properties to characterize the desired configuration, which are enforced through explicit geometric constraints between neighboring agents. Classical approaches include distance-based schemes (\cite{Krick2009,Queiroz2019}), and bearing-based schemes (\cite{ZHAO2019_CSM}).

A central challenge lies in balancing sparse information exchange with guaranteed convergence to the desired configuration. To address this challenge, distance and bearing-based schemes leverage results from rigidity theory, relying on minimal infinitesimal rigidity (MIR) as a crucial architectural requirement to guarantee convergence to a desired shape (\cite{Krick2009, ZHAO2019_CSM}). In $\mathbb{R}^2$, MIR requires  $2n-3$ communication links to achieve a unique (local) formation of $n$ agents (up to translations, rotations, and flip ambiguities). However, such rigidity-based architectures typically require a large number of inter-agent constraints, increasing sensing and communication complexity.

A parallel line of research has explored augmented Laplacian formulations, where the inter-agent geometric relations are encoded via structured weights (\cite{Trinh_AUT2018}), including complex-Laplacian approaches (\cite{deMarina2020, Zhou2025}) in which the scalar entries are replaced with complex numbers, naturally capturing rotations, translations, and scalings. \textit{Matrix-weighted Laplacian} formulations, where the weights are structured matrices encoding inter-agent relations, offer another approach. Examples include projection-matrix weights seen in bearing-based control (\cite{ZHAO2019_CSM}), and matrix-structured weights that mimic the role of complex weights (\cite{Fan2025}).

Within this research field, spatial \textit{symmetries} offer a potential alternative to rigidity-based approaches (\cite{Zelazo2025forced,Martinez2025}). In many practical formations, the geometric relationships between agents naturally appear as rotations or reflections (about a common centroid or axis), often arising from sensing coverage or communication requirements. By exploiting these relationships, traditional geometric constraints (e.g., distances or bearings) can be replaced with point group relations, significantly reducing the number of required inter-agent communication links. Recent results (\cite{Martinez2025}) demonstrate the potential of this approach as a structured matrix-weighted Laplacian, showing that rotational symmetry constraints can drive the system to a unique symmetric configuration while using only $n-1$ inter-agent communication links in $\R^2$, matching the minimal connectivity required between the agents.

Planar formations frequently exhibit dihedral symmetries that include both rotations and reflections. This naturally motivates studying reflection relations as a complementary approach in symmetry-based formation control. While rotational relations naturally encode cyclic geometry, reflectional relations pose a more challenging problem. Enforcing mirror constraints alone may produce large families of admissible configurations, leading to symmetric flexes. As a consequence, designing a distributed formation control law based solely on reflections to achieve a unique symmetric shape is non-intuitive and remains an open problem.

This work extends the symmetry-based framework in $\R^2$ by demonstrating the potential of reflection constraints. We show that although inter-agent reflection constraints alone may lead to ambiguous configurations, introducing a leader-follower architecture resolves this ambiguity and suffices to solve the symmetry-based formation control problem using \textit{only} $n-1$ communication links. We provide a stability and convergence analysis of the resulting control law. In addition, we propose an extension of the proposed strategy to allow the multi-agent system to achieve and maintain the desired configuration while undergoing coordinated translations, rotations, and scalings along a time-varying reference trajectory known to all agents.
\paragraph*{Notations} 
A (finite, simple) graph $\G=(\V,\E)$ consists of two non-empty sets, the set of vertices (nodes) $\V=\{1,...,n\}$,  and the set of edges $\E=\{e_1,\dots,e_m\}\subseteq \V \times \V$. Throughout, $\G$ is assumed to be undirected. The notation $ij\in\E$ indicates that agent $i$ can receive information from its neighboring agent $j$, and vice versa. The graph Laplacian of $\G$ is defined as $L(\G)=EE^T$, where $E\in\R^{n\times m}$ is the incidence matrix (using an arbitrary orientation of the edges) with $[E]_{ij}=1$ if the edge $e_j$ leaves
vertex $i$, $[E]_{ij}=-1$ if the edge $e_j$ enters the vertex $i$, and $[E]_{ij}=0$, otherwise. A path $\mathrm{path}(i,j)$ is a sequence of
distinct vertices from $i$ to $j$ such that consecutive vertices are adjacent.
Let $I_n\in\mathbb{R}^{n\times n}$ denote the identity matrix, $\mathds{1}_n \in \mathbb{R}^n$ denote the all-one column vector of dimension $n$, and ${\mathrm e}_i$ denote the $i$-th standard basis vector in $\R^n$. The Kronecker product is denoted by $\otimes$.
\section{Preliminaries}\label{sec.preliminaries}
We provide in this section an overview of the key notions from graph theory and group theory that will be used throughout the paper.
\subsection{Symmetry in Graphs}
Group theory provides a fundamental mathematical framework for describing symmetries. We recall the basic definitions used throughout this work.
\begin{definition}
A \emph{group} is a set $\Gamma$ equipped with a binary operation $\circ$ satisfying:
\begin{itemize}
    \item[i)] \emph{Closure:} For all $a,b \in \Gamma$, the composition $a \circ b$ is also in $\Gamma$.
    \item[ii)] \emph{Associativity:} $(a \circ b) \circ c = a \circ (b \circ c)$ for all $a,b,c \in \Gamma$.
    \item[iii)] \emph{Identity:} There exists an element $\mathrm{id} \in \Gamma$ such that $a \circ \mathrm{id} = \mathrm{id} \circ a = a$ for all $a \in \Gamma$.
    \item[iv)] \emph{Inverses:} For each $a \in \Gamma$, there exists an inverse $a^{-1} \in \Gamma$ such that $a \circ a^{-1} = a^{-1} \circ a = \mathrm{id}$.
\end{itemize}
The \emph{order} of a group is the number of its elements. A subset $B\subseteq\Gamma$ that is itself a group under $\circ$, is called a \emph{subgroup}.
\end{definition}

\begin{definition}
Let $\G = (\V, \E)$ be a finite, simple graph. An \emph{automorphism} of $\G$ is a permutation $\psi : \V \to \V$ such that $
    uv \in \E \; \Leftrightarrow \; \psi(u)\psi(v) \in \E.
$
\end{definition}
One can express every permutation as a composition of disjoint cycles of the permutation. A cycle is a successive action of the permutation that sends a vertex back to itself, i.e., $i\to \psi(i) \to \psi(\psi(i)) \to \cdots \to \psi^k(i)=i$, where $\psi^k = \underbrace{\psi \circ \cdots \circ \psi}_{k \text{ times }}$.  {Such a cycle is compactly written using the \emph{cycle notation}, denoted by $(i\,\psi(i)\,\cdots \psi^{k-1}(i))$.} The integer $k$ is the \emph{length} of the cycle.
In the context of graphs, the set of all automorphisms of $\G$ forms the group $\Aut(\G)$ under composition.  This group captures the \emph{symmetries} of the graph: each element of \(\Aut(\G)\) is a relabeling of the vertices that preserves the adjacency structure.

Often, one works not with the full automorphism group but with a chosen subgroup
$\Gamma \subseteq \Aut(\G)$.
In this case, we say informally that “$\Gamma$ acts as symmetries of the graph,”
or simply that $\G$ is $\Gamma$-symmetric.
Different choices of subgroups capture distinct types of structural symmetries. Below we consider an example of a simple cycle graph to demonstrate how rotational and reflectional symmetries arise naturally and how selecting particular subgroups encodes the sufficient symmetrical structure of the graph.

\begin{example} \label{ex:aut4}
Fig.~\ref{fig:c4_ex1} shows the cycle graph $C_4$. We can directly identify all elements of its automorphism group $\Aut(C_4)$. 
Consider clockwise rotations of $90^\circ$, $180^\circ$, and $270^\circ$, which yield the automorphisms (in cycle notation) $\psi_1=(1\,2\,3\,4)$, $\psi_2=\psi_1^2=(1\,3)(2\,4)$ and  $\psi_3=\psi_1^3=(1\,4\,3\,2)$, respectively.
Additional permutations can be found by considering reflections. The reflection about the vertical red axis gives the permutation $\psi_4=(1\,2)(4\,3)$, while the reflection about the horizontal blue axis yields $\psi_5=(1\,4)(2\,3)$. Additionally, the diagonal reflections about the green and orange axes yield $\psi_6=(1)(2\,4)(3)$ and $\psi_7=(2)(1\,3)(4)$, respectively. Thus, we have that $\Aut(C_4)=\{\mathrm{id},\psi_1,\ldots,\psi_7\}$ contains 8 automorphisms.
\begin{figure} [!h]
    \begin{center}   
    \vspace{-0cm}
    \includegraphics[width=0.2\linewidth]{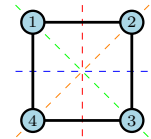}
    \end{center}
\vspace{-0.1cm}
\caption{Cycle graph $C_4$, with $8$ automorphisms in $\Aut(\G)$.}\label{fig:c4_ex1}
\end{figure}
\vspace{-0.2cm}

We may choose $\Gamma=\{\mathrm{id}, \psi_1,\psi_1^2,\,\psi_1^3\}$, which corresponds to the subgroup of rotational automorphisms of $C_4$. Under this choice, $C_4$ is considered as $\Gamma$-symmetric where any vertex can be mapped to any other under the rotation actions of $\Gamma$. 
\end{example}

\subsection{Symmetry in frameworks}\label{sec:symmetry_fwks}
When dealing with formation control problems, the embedding of symmetric graphs in Euclidean space is of particular interest. In this direction, we now consider \textit{symmetry of frameworks} (\cite{Bernd2017sym}). A framework in $\mathbb{R}^2$ is defined as the pair $(\G,p)$, where $p: \V\rightarrow\mathbb{R}^2$ assigns each vertex $i\in\V$ a position $p_i\in\R^2$ representing the physical location of the agents. We denote the configuration of the entire network as the stacked vector $p = \begin{bmatrix}p_1^T &\cdots & p_n^T\end{bmatrix}^T$. 

We associate each automorphism of the graph with a corresponding Euclidean isometry (a rotation or reflection) through an appropriate group representation.

\begin{definition}\label{def:tau-gamma}
Let $\Gamma \subseteq \mathrm{Aut}(\G)$, and let $\tau:\Gamma\rightarrow O(\mathbb{R}^2)$ be a group homomorphism giving an orthogonal representation of $\Gamma$ on $\mathbb{R}^2$. A framework $(\G,p)$, with $p\in \mathbb{R}^2$, is called \emph{$\tau(\Gamma)$-symmetric} if 
\begin{equation}\label{eq:symfwk}
\tau(\gamma) p_i=p_{\gamma (i)} \quad \forall \gamma\in \Gamma ,\quad i\in \V.
\end{equation}
\end{definition}
%
We use the standard Schoenflies notation for point groups (\cite{atk70,alt94}). In $\R^2$, the possible point groups are
\begin{itemize}
    \item[i)] \emph{Rotational groups} $\C_n$: the cyclic group generated by an $n$-fold rotation $c_n$ about the origin.
    \item[ii)] \emph{Reflection groups} $\C_s$: a point group consisting of a single mirror reflection across a fixed line in~$\mathbb{R}^2$.
    \item[iii)] \emph{Dihedral groups} $\C_{nv}$: the point group of order $2n$, generated by $\C_n$ together with $n$ distinct reflections $\{\sigma_1,\dots,\sigma_n\}$ whose mirror lines are spaced by angles of $2\pi/n$.
\end{itemize}
The rotational subgroup $\C_n$ and each reflection subgroup 
$\C_s^{(k)} := \{e,\sigma_k\}$ embed naturally in $\C_{nv}$, i.e.,
$\C_n \subset \C_{nv}$, and $\C_s^{(k)} \subset \C_{nv}$ (for $k=1,\dots,n$).  The reflection automorphisms ($\C_s^{(k)}$) of the cycle graph $C_n$ naturally divide into two geometrically distinct classes.  We introduce the notation $\mathscr{F}_s$  and  $\mathscr{S}_s$ to distinguish between them:
\begin{itemize}
    \item[i)] \emph{Free reflections}:
    $$
        \mathscr{F}_s :=
        \{\sigma \in \mathcal{C}_{nv} : 
        \sigma(i) \neq i \text{ for all } i\in \V\}.
    $$
    
    \item[ii)] \emph{Self reflections}:
    $$
        \mathscr{S}_s :=
        \{\sigma \in \mathcal{C}_{nv} :
        \sigma(i) = i \text{ for some } i\in \V\}.
    $$
\end{itemize}

These sets form a disjoint partition of the reflections of
$\mathcal{C}_{nv}$:
$$
\mathscr{R}_s:=\{\sigma_1,\dots,\sigma_n\} = 
\mathscr{F}_s \;\cup\; \mathscr{S}_s.
$$
Free reflections encode mirror relations between distinct agents,
while self reflections constrain agents to lie on specific mirror axes.
Both types are required to realize the full dihedral symmetry $\C_{nv}$.

By considering the graphs in Fig. \ref{fig:sym_fw_Cn} as planar frameworks $(C_n,p)$, their geometric symmetry depends entirely on the embedding. Depending on the embedding, the framework may exhibit the full dihedral symmetry $\C_{4v}$ (Fig. \ref{fig:sym_fw_Cn}(d)), only the rotational symmetry $\C_4$ (Fig. \ref{fig:sym_fw_Cn}(c)), a single reflection symmetry $\C_s$ (Fig. \ref{fig:sym_fw_Cn}(b)) or no nontrivial symmetry at all (Fig. \ref{fig:sym_fw_Cn}(a)).

As outlined in Definition \ref{def:tau-gamma}, the geometric action of the point groups can be represented through a homomorphism $\tau : \Gamma \to O(\mathbb{R}^2)$. For any two vertices $i,j\in\V$, we denote by $\gamma_{ij}\in\C_n$ the rotational group element satisfying \eqref{eq:symfwk}, i.e., the automorphism that maps $i\mapsto j$. The representation of the element $\tau(\gamma_{ij})$ is given as the standard planar rotation by an angle $\theta=2\pi/n$
$$\tau(\gamma_{ij})=R(\theta)=\begin{bmatrix}
    \cos(\theta) & -\sin(\theta) \\ \sin(\theta) & \cos(\theta)
\end{bmatrix},$$
and the inverse relation satisfies $\tau(\gamma_{ji})=R(-\theta)=R(\theta)^T=\tau(\gamma_{ij})^T$.
We also use the notation $\gamma_{ij} \in \fr_s$ for reflection-induced automorphisms mapping $i\mapsto j$. Let $\mathcal{L}_{ij}$ denote the corresponding mirror line, and  $\hat{n}_{ij}$ its unit normal vector.
The representation of $\gamma_{ij}$ is given by the Householder matrix
$$
\tau(\gamma_{ij})=H(\hat{n}_{ij})=I_2 - 2\, \hat{n}_{ij} \hat{n}_{ij}^\top.
$$
This matrix satisfies $\tau(\gamma_{ij})^2 = I_2$, $\tau(\gamma_{ij})=\tau({\gamma_{ji}})$, and
$\det(\tau(\gamma_{ij})) = -1$, as expected for a reflection. Similarly, for a self reflection we denote $\gamma_i\in\sr_s$, with mirror line $\mathcal{L}_i$ and normal vector $\hat{n}_i$.

\section{Symmetry-Based \\Formation Control for $C_n$}\label{sec.formation}

In this section we introduce the symmetry-based formation control problem, specialized for the cycle graph $C_n$.  We choose the cycle graph since it is the simplest nontrivial graph whose automorphism group contains the full dihedral symmetry $\C_{nv}$.  Thus, it can serve as a preliminary study to understand how to design symmetry-constrained control strategies.  The control objective, therefore, is to design distributed controllers that drive the agents to a formation exhibiting the \emph{full dihedral symmetry, $\C_{nv}$}, consisting of both rotational and reflectional relations.

Towards this goal, we proceed in stages.  First, we briefly review our previous work on rotational-symmetry formation control, showing that it solves the problem for the rotational subgroup $\C_n$.  Next, we propose an initial control strategy that enforces inter-agent reflectional relations for free reflections only.  
Finally, we introduce an augmented reflection-based control strategy that is able to stabilize the formation to the full $\C_{nv}$ symmetry.

\subsection{Problem Setup}
We consider a team of $n$ agents modeled by the integrator dynamics
\begin{align}\label{int-dynamics}
\dot p_i(t) = u_i(t), \quad i \in \{1, \ldots, n\},
\end{align}
where $p_i(t) \in \mathbb{R}^2$ is the position of agent $i$ and $u_i(t) \in \mathbb{R}^2$ is its control to be designed. The formation control objective is to steer the state to a configuration that exhibits the full dihedral symmetry of $C_n$, that is, to a $\tau(\C_{nv})$-symmetric configuration. Formally, we want to render the set 
$$\mathcal F = \{p \in \mathbb{R}^{2n} \, |\, \tau(\gamma)p_i=p_{\gamma(i)} \text{ for all } \gamma \in \C_{nv} \text{ and } i \in \V\},$$
asymptotically stable. Then, the objective is to satisfy
\begin{align}\label{sym_problem}
    \lim_{t \to \infty} \big\| p_i(t) - \tau(\gamma_{ji})\, p_j(t) \big\| = 0,
\end{align}
for all $\gamma_{ji}\in\C_n\cup\mathscr{F}_s$, and
\begin{align}\label{sref_problem}
   \lim_{t \to \infty} \big\| p_i(t)-\tau(\gamma_i)p_i(t) \big\| = 0,
\end{align}
for $\gamma_i \in \mathscr{S}_s$.

\begin{figure}[!t]
\begin{center}
\includegraphics[width=0.95\linewidth]{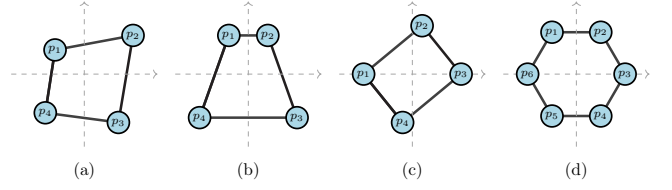}
    \end{center}
\caption{Symmetric frameworks with $C_n$ as the underlying graph. (a) exhibits no nontrivial symmetry, (b) has a reflection symmetry ($\C_s$) about the $y$-axis, (c) has a $180^\circ$ rotation symmetry ($\C_n$), and (d) is $\mathcal{C}_{6v}$-symmetric.}
\label{fig:sym_fw_Cn}
\end{figure}

We define an \emph{interaction graph} $\G_I = (\V,\E_I)$ to specify which 
agents are able to exchange information. This graph is defined as a spanning tree subgraph of $C_n$ which ensures the minimal connectivity required in the MAS. For example, for the $\mathcal{C}_{4v}$-symmetric frameworks seen in Fig. \ref{fig:sym_fw_Cn}(b) and (c), the 
edge set $\E_I = \{12,23,34\}$ satisfies the connectivity
requirement.

\subsection{Rotational Symmetry Formation Control}
We begin by reviewing our previous work (\cite{Martinez2025}) which provided a solution to the symmetry-based formation control problem for the rotational subgroup $\C_n$. The first step is to define a \textit{rotational symmetry-forcing potential} over the edges in the interaction graph,
\textcolor{black}{\begin{align}\label{sympotential1}
F_{\C_n}(p(t)) = \frac{1}{2} \sum_{ij\in \mathcal{E}_I}  \|p_i(t) - \tau(\gamma_{ji} )p_j(t)\|^2,
\end{align}}%
where $\gamma_{ij} \in \C_n$. We then implement a control defined by the gradient dynamical system
\begin{align}\label{symform_acquire}u(t) = -\nabla F_{\C_n}(p(t)),
\end{align}
and examine the expression of the closed-loop
dynamics for each agent. For an agent $i$ the dynamics are
\begin{align}\label{ea_dyn}
    \dot{p}_i(t)&= \sum_{{ij\in\mathcal{E}_I}}(\tau({\gamma_{ji}})p_j(t)-p_i(t)). 
\end{align}
Expressing the dynamics in state-space form yields the closed-loop system
\begin{align}\label{ctrl_r}
    \dot{p}(t) = - Q(\C_n)p(t),
\end{align}
where $Q(\C_n)\in\R^{2n\times 2n}$ is the resulting \textit{symmetry-constraining} matrix-weighted Laplacian for the graph $\G_I$, with block entries
\textcolor{black}{\begin{equation}\label{eq.Q} [Q(\C_n)]_{ij} = \begin{cases}
                d(i)I_2, & i=j, \, i \in \V \\
                -\tau(\gamma_{ji}), & ij\in \mathcal{E}_I \\
                0, & \text{o.w.}
            \end{cases},\end{equation}
where $d(i)$ denotes the degree of vertex $i$ in $\G_I$}, and $\gamma_{ij} \in \C_n$. 
The matrix $Q(\C_n)$ has the same structural form used in matrix-weighted Laplacian methods (e.g., \cite{ZHAO2019_CSM}).
 As a matrix-weighted Laplacian, it can be expressed as the matrix product $E(\C_n)E(\C_n)^T$, where $E(\C_n)\in\mathbb{R}^{2|\V| \times 2|\E_I|}$ results in a matrix-weighted incidence matrix structure similar to the scalar incidence matrix. For an edge $k=ij$, one
has $[E(\C_n)]_{ik}=I$ if the edge $k$ leaves vertex $i$, $[E(\C_n)]_{ik}=-\tau(\gamma_{ji}) $ if the edge $k$ enters the vertex $i$, and $[E(\C_n)]_{ik}=0$ otherwise. 
\begin{theorem}[\cite{Martinez2025}]\label{th_r}
    Consider a MAS consisting of $n$ integrator agents \eqref{int-dynamics}, whose interaction topology is defined  by a spanning tree graph $\G_I \subset C_n$, and let
    $$\mathcal{F}_{\C_n}=\{p\in\R^{2n}|E(\C_n)^Tp=0,\;\;i\in \V\}.$$
    Then, for any initial condition $p(0) \in \mathbb{R}^{2n}$, 
    the control 
    \begin{align*}
        u(t)=-Q(\C_n)p(t)
    \end{align*}
    renders the set $\mathcal{F}_{\C_n}$ exponentially stable.
\end{theorem}

Theorem~\ref{th_r} establishes that trajectories of the closed-loop system converge exponentially to the set $\mathcal{F}_{\C_n}$, which is precisely the set of configurations for which every enforced rotational relation is satisfied. Equivalently, $\mathcal{F}_{\C_n}=\Null\,\big(Q(\C_n)\big)$, implying that a steady-state configuration $p^\star$ satisfies
$$
    E(\C_n)^T p^\star = 0.
$$
Therefore, the control law \eqref{ctrl_r} steers the agents to a unique $\C_{n}$-symmetric configuration (up to trivial rotations and scalings). 

\begin{example}\label{ex:c6_ex_r}
Consider a MAS consisting of $n=6$ agents, tasked with attaining a $\C_{6}$-symmetric configuration. Fig. \ref{fig:c6_combined}(a) illustrates the underlying $C_6$ graph, where the dashed edge is removed to obtain the chosen interaction graph $\G_I$, defined by the edge set $\E_I = \{12, 23, 34, 45, 56 \}$. 
Note that by using a rigidity-theoretic approach to solve the formation control problem, $9$ edges would be required (see \cite{Krick2009,ZHAO2019_CSM}). 
\begin{figure}[ht]
    \centering

    \begin{subfigure}{0.35\linewidth}
        \centering
        \includegraphics[width=0.7\linewidth]{figures/c6_sim_gr.tex}
        \vspace{0.8cm}
        \hspace{-0.2cm}
        \caption{The cycle graph $C_6$\\.}
        \label{fig:sim1_tg}
    
    \end{subfigure}
    \hfill
    \begin{subfigure}{0.6\linewidth}
        \centering
        \includegraphics[width=0.8\linewidth]{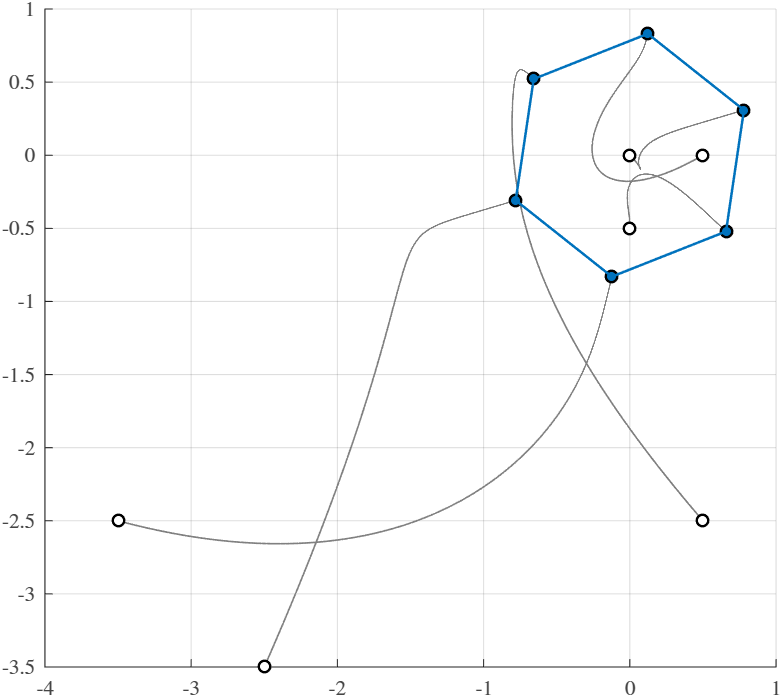}
        \caption{Agents converge to a $\tau(\C_n)$-symmetric formation.}
        \label{fig:c6_sim_r}
    \end{subfigure}

    \caption{Graph structure and simulation results for the $\C_n$-symmetric  formation control problem in Example~\ref{ex:c6_ex_r}.}
    \label{fig:c6_combined}
\end{figure}

The actions in $\tau(\C_6)$ correspond to rotations of $\pi/3$ about the origin. The respective matrix $Q(\C_6) \in \mathbb{R}^{12\times 12}$ corresponds to the interaction graph $\E_I$ and has the form
$$
Q(\C_6) = 
\begin{bmatrix}
I_2      & -\tau_r & 0       & \cdots  & 0      \\
-\tau_r^T& 2I_2    & -\tau_r & \ddots  & \vdots \\
0        & \ddots  & \ddots  & \ddots  & 0      \\
\vdots   & \ddots  & -\tau_r^T& 2I_2   & -\tau_r \\
0        & \cdots  & 0       & -\tau_r^T & I_2
\end{bmatrix},
\quad
\tau_r = R(\pi/3).
$$

Fig. \ref{fig:c6_combined}(b) shows the trajectories and final configuration obtained under the control law \eqref{ctrl_r}.

\end{example}

In the setting of Example~\ref{ex:c6_ex_r}, observe that under control law \eqref{ctrl_r}, the agents converge to the expected $\C_{6}$-symmetric configuration, i.e., a rotated realization of the framework shown in Fig.~\ref{fig:sym_fw_Cn}(d). However, while this enforces the rotational components of the symmetry-forcing scheme, it does not enforce the full dihedral symmetry of the target framework (undesired mirror symmetries).

\subsection{Free Reflectional Symmetry Formation Control}

Having characterized the rotational case, we now consider a similar strategy to solve the problem for reflectional symmetries. Following the same procedure as for the rotation case, we begin by focusing solely on the free (inter-agent) reflection relationships encoded by the edges. We define the \emph{reflectional symmetry-forcing potential}
\begin{align}\label{sympotential_reflection}
F_{\mathscr{F}_s}(p(t)) = \frac{1}{2} \sum_{ij\in \mathcal{E}_I}  \|p_i(t) - \tau(\gamma_{ji} )p_j(t)\|^2,
\end{align}
where $\gamma_{ij} \in \mathscr{F}_s$.  

Implementing the negative gradient flow, $$u(t) = -\nabla F_{\mathscr{F}_s}(p(t)),$$ yields the reflection-based control law
\begin{align}\label{ctrl_s}
    \dot{p}(t) = -Q(\mathscr{F}_s)\,p(t),
\end{align}
where $Q(\mathscr{F}_s)$ is the matrix-weighted Laplacian associated with the free reflections set $\mathscr{F}_s$.  Note that this matrix has the same structure as $Q(\C_n)$ in \eqref{eq.Q}. 

We can note some immediate properties of $Q(\mathscr{F}_s)$. First, recall that $Q(\mathscr{F}_s)=E(\mathscr{F}_s)E(\mathscr{F}_s)^T$, and hence is symmetric and PSD.

Consequently,
$$
    p \in \Null\big(Q(\mathscr{F}_s)\big)
    \quad \Longleftrightarrow \quad
    E(\mathscr{F}_s)^T p = 0.
$$
We now state a result on the dimension of the null-space of $Q(\mathscr{F}_s)$.
\begin{proposition}\label{prop:Qs_rank}
The free reflection-based matrix-weighted Laplacian $Q(\mathscr{F}_s)$ satisfies $\dim \Null(Q(\mathscr{F}_s)) = 2.$
\end{proposition}
\begin{proof}
First, note that the condition $E(\mathscr{F}_s)^Tp=0$ encodes all reflection relations $p_i=\tau(\gamma_{ji})p_j$ associated with the edges of $\G_I$. 
Along any  path between two nodes $i$ and $j$, the symmetry constraints between neighboring agents yield the recursive relations 
$$
p_{v_{j+1}} = \tau(\gamma_{v_j\,v_{j+1}})\, p_{v_j}, \qquad j=0,\dots,k-1,
$$ 
where $v_0=i$, $v_k=j$, and $v_\ell v_{\ell+1} \in \E_I$.
Composing these relations yields the chain 
$$
p_j = \underbrace{\tau(\gamma_{v_{k-1}\, v_k})\tau(\gamma_{v_{k-2}\, v_{k-1}})\cdots\tau(\gamma_{v_0, v_1}) }_{S_j \in O(2)}\, p_i,
$$
such that 
$$
   p_j = S_j p_i,\quad S_j = \tau(\gamma_{ji}) \in O(2),
$$ 
for all $j=1,\dots,n$, with $S_i = I_2$. 
Hence, every configuration in the nullspace can be written as
$$   
p =
    \begin{bmatrix}
        p_1^T & \cdots & p_n^T
    \end{bmatrix}^T
    = \begin{bmatrix}
        (S_1p_\l)^T & \cdots & (S_np_\l)^T
    \end{bmatrix}^T,
$$
for an arbitrary vertex $\l \in \V$.
This shows
$$
    \Null\big(Q(\mathscr{F}_s)\big)
    = \big\{\,p_\l \in \R^2 \big| [(S_1 p_\l)^T \; \cdots \; (S_np_\l)^T]^T \big\},
$$
which is a 2-dimensional subspace parameterized by $p_\l \in \R^2$. Therefore, this implies $\dim \Null(Q(\mathscr{F}_s)) = 2$.
\end{proof}

\begin{theorem}\label{th_s}
    Consider a MAS consisting of $n$ integrator agents \eqref{int-dynamics}, whose interaction topology is defined  by a spanning tree graph $\G_I\subset C_n$, and let
    $$\mathcal{F}_{\mathscr{F}_s}=\{p\in\R^{2n}|E(\mathscr{F}_s)^Tp=0,\;\;i\in \V\}.$$
    Then, for any initial condition $p(0) \in \mathbb{R}^{2n}$,
    the control 
    \begin{align*}
        u(t)=-Q(\mathscr{F}_s)p(t)
    \end{align*}
    renders the set $\mathcal{F}_{\mathscr{F}_s}$ exponentially stable.
\end{theorem}

\begin{proof}

Since $Q(\mathscr{F}_s)$ is symmetric, PSD, and satisfies $\dim \Null(Q(\mathscr{F}_s)) = 2$, there exists an orthonormal matrix $V$ and diagonal matrix $\Lambda = \operatorname{diag}(0,0,\lambda_{3},\dots,\lambda_{2n})$ with $\lambda_k > 0 $ for all $k\geq 3$, such that
$$
    Q(\mathscr{F}_s) = V\Lambda V^T.
$$
Let $V = \begin{bmatrix} V_0 & V_+\end{bmatrix}$, where $V_0 \in \mathbb{R}^{2n \times 2}$ contains an orthonormal basis for $\Null\big(Q(\mathscr{F}_s)\big)$.  Similarly, partition $\Lambda = \operatorname{diag}(0_2,\Lambda_+)$ so $0_2$ is the $2\times 2$ matrix of zeros.
The solution of \eqref{ctrl_s} is then
$$
    p(t) = e^{-Q(\mathscr{F}_s)t}\,p(0)= V_0V_0^Tp(0)+ V_+e^{-\Lambda_+ t} V_+^T p(0).
$$
As $t\to\infty$, we obtain
$$
    \lim_{t\to\infty} p(t)
    = V_0 V_0^T p(0).
$$
Thus,  the trajectories converge to the subspace
$$
    \mathcal{F}_{\mathscr{F}_s}
    = \Null\big(Q(\mathscr{F}_s)\big)
    = \Big\{q \in \mathbb{R}^2 \big|[(S_1 q)^T \; \cdots \; (S_n q)^T]^T \Big\},
$$
rendering $\mathcal{F}_{\mathscr{F}_s}$ exponentially stable, as claimed.
\end{proof}

We conclude that the subspace $\mathcal{F}_{\mathscr{F}_s}$ does not enforce the self-symmetry condition \eqref{sref_problem} since this condition applies if and only if each agent $i$ lies on its corresponding mirror line $\mathcal{L}_i$.  To show this explicitly, recall that $p \in \mathcal{F}_{\mathscr{F}_s}$ implies $p_i = S_i p_j$ for some $p_j \in \mathbb{R}^2$ and all $i\in\V$. Therefore, $\mathcal{F}_{\mathscr{F}_s}$ also admits configurations satisfying $p_j \notin \mathcal{L}_j$. Hence, the reflection-based control law \eqref{ctrl_s} alone may drive the system to a "symmetric flexible" configuration satisfying all  pairwise reflection constraints that does not coincide with the intended $\C_{nv}$-symmetric target formation.

\begin{example}\label{ex:c6_ex_s}
Consider the same setup as in Example \ref{ex:c6_ex_r}, however, the agents are now tasked with attaining a symmetric configuration under inter-agent free reflection constraints.
The actions $\tau(\mathscr{F}_s)$ include the vertical mirror $\mathcal{L}_{12}$, which reflects agent pairs $(1,2)$ and $(4,5)$, as well as two diagonal mirrors: $\mathcal{L}_{23}$, reflecting $(2,3)$ and $(5,6)$, and $\mathcal{L}_{34}$, reflecting $(3,4)$ and $(1,6)$. The actions are represented by using the Householder reflection $\tau(\mathcal{L}_{ij}) = H(\hat{n}_{ij})$. The matrix-weighted Laplacian $Q(\mathscr{F}_s)$ has the same structure the one in Example \ref{ex:c6_ex_r}.
Fig. \ref{fig:c6_sim_s} shows the trajectories and final configuration obtained under the control law \eqref{ctrl_s}.

\begin{figure}[ht]
    \centering
        \centering
        \includegraphics[width=0.48\linewidth]{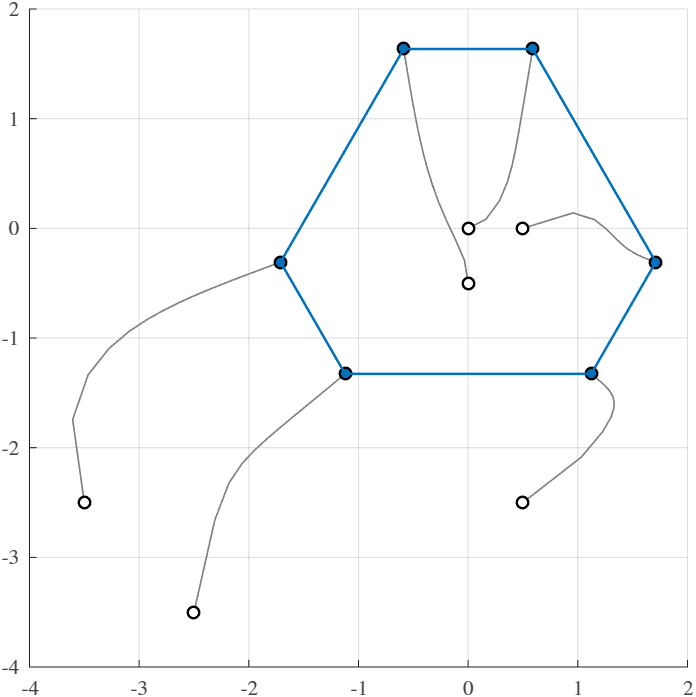}
    \caption{Simulation results for Example~\ref{ex:c6_ex_s}.}
    \label{fig:c6_sim_s}
\end{figure}

\end{example}

In Example~\ref{ex:c6_ex_s}, observe that control law \eqref{ctrl_s} drives the agents to a configuration that satisfies the prescribed pairwise reflection relations but does not enforce the full dihedral symmetry of the desired $C_{6v}$-symmetric configuration. 

\subsection{Full $\mathcal{C}_{nv}$ Symmetry Formation Control}

We now propose an augmentation of the reflection-based control law so that the closed-loop dynamics enforce all symmetry relations in $\gamma\in\mathscr{R}_s=\mathscr{F}_s \cup \mathscr{S}_s$. As a key result in this work we show that enforcing all free symmetries in addition to a single designated agent $\l$ anchored to its prescribed mirror line $\mathcal{L}_\l$ suffices to satisfy all relations $\gamma\in\C_{nv}$ of a $\C_{nv}$-symmetric configuration.

We augment the potential~\eqref{sympotential1} with the term
\begin{align}\label{eq:Fsr}
    F_{\l}=\frac{1}{4}  \|p_\l(t) - \tau(\gamma_\l )p_\l(t)\|^2,
\end{align}
for $\gamma_\l \in \mathscr{S}_s$.
We now define the augmented gradient control law
\begin{align}\label{ctrl_Faug}
    u(t) =& -\nabla (F(p(t))+F_\l(p(t)))\nonumber\\
    =& -Q(\mathscr{F}_s)p(t)-\diag({\mathrm e}_\l)\otimes (I_2 - \tau(\gamma_\l))p(t).
\end{align}

This yields the following augmented matrix-weighted Laplacian
\begin{equation}\label{eq:refl_aug}
    \mathcal{Q}(\mathscr{R}_s)=
        Q(\mathscr{F}_s) + \diag({\mathrm e}_\l)\otimes (I_2 - \tau(\gamma_\l)),
\end{equation}
where $Q(\mathscr{F}_s)$ is the matrix-weighted Laplacian associated with the free reflection set $\mathscr{F}_s$.

Intuitively, since the interaction graph is a spanning tree, the single constraint $\gamma_\l\in\sr_s$ propagates through the reflection relations to all agents.

\begin{lemma}
    Let $\G_I \subset C_n$ be a spanning tree.  For each edge $ij \in \E_I$, let $\gamma_{ji} \in \mathscr{F}_s$ be the free reflection between agents $i$ and $j$.  Let $\l \in \V$ be a designated node constrained to the mirror line $\mathcal L_\l \subset \R^2$ corresponding to a self-reflection $\gamma_\l \in \mathscr{S}_s$. Then there exists a unique assignment of mirror lines $\mathcal L_i$ for each $i\in \V$, such that for every edge $ij\in \E_I$ the mirror lines satisfy $\mathcal L_j = \tau(\gamma_{ji})\mathcal L_i$.  In particular, the orientation of every agents' mirror axis $\mathcal L_i$ is uniquily determined by $\mathcal L_\l$ and the inter-agent reflection symmetries.
\end{lemma}
\begin{proof}
Since $\G_I$ is a spanning tree, there is a unique path from $\l$ to any node $i$.  Along this path define
$$
S_i := \tau(\gamma_{v_k v_{k-1}})\cdots \tau(\gamma_{v_1 v_0}) \in O(2),
$$
with $S_\ell = I_2$, $v_0=\l$ and $v_k=i$.  
We now \emph{define} the mirror line of agent $i$ by
$$
\mathcal L_i := S_i \mathcal L_\l.
$$
This assignment satisfies the required inter-agent relation because, if
$ij$ is an edge and $j$ is farther from $\l$ than $i$, then by
construction $S_j = \tau(\gamma_{j i})\, S_i$, and therefore  
$$
\mathcal L_j = S_j \mathcal L_\l = \tau(\gamma_{j i}) \mathcal L_i.
$$
Thus a valid assignment exists.

To establish uniqueness, let $\{\mathcal L'_i\}_{i=1}^{|\V|}$ be any other assignment with $\mathcal L'_\l = \mathcal L_\l$ and  
$\mathcal L'_j = \tau(\gamma_{j i}) \mathcal L'_i$ for all edges $ij\in E_I$.
Since every agent is reached from $\l$ by a unique path, an induction
along that path gives
$$
\mathcal L'_i= \tau(\gamma_{i v_{k-1}})\cdots \tau(\gamma_{v_1 v_0}) \mathcal L'_\l= S_i \mathcal L_\l= \mathcal L_i.
$$
Hence $\mathcal L'_i = \mathcal L_i$ for all $i$, proving uniqueness.
\end{proof}

We now consider the augmented reflection-based control law
\begin{align}\label{ctrl_aug}
    \dot{p}(t) = -\qq(\mathscr{R}_s)\,p(t).
\end{align}

The properties of the matrix $\mathcal{Q}(\mathscr{R}_s)$ are of interest. First, note that $I_2-\tau(\gamma_\l)=2\hat{n}_\l{\hat{n}_\l}^T$. Since 
$$
\mathcal{Q}(\mathscr{R}_s)=E(\mathscr{F}_s)E(\mathscr{F}_s)^T + \diag({\mathrm e}_\l)\otimes2\hat{n}_\l{\hat{n}_\l}^T,$$
for any $p\in\R^{2n}$ we obtain
\begin{align*}
p^T\mathcal{Q}(\mathscr{R}_s)p=&p^T\left(E(\mathscr{F}_s)E(\mathscr{F}_s)^T+\diag({\mathrm e}_\l)\otimes2\hat{n}_\l{\hat{n}_\l}^T\right)p\\=&\|E(\mathscr{F}_s)^Tp\|^2+2|\hat{n}_\l^Tp_\l|^2\geq 0.    
\end{align*} 
Hence, $\mathcal{Q}(\mathscr{R}_s)$ is PSD and symmetric. Additionally, for any $p\in\R^{2n}$ we have
$$
p^T\mathcal{Q}(\mathscr{R}_s)p=0 \quad \Longleftrightarrow \quad \begin{cases} E(\mathscr{F}_s)^Tp&=0 \\
 (I_2-\tau(\gamma_\l))p_\l &= 0\end{cases}.
$$

\begin{proposition}\label{prop:Qs_aug_rank}
The augmented reflection-based matrix-weighted Laplacian $\qq(\mathscr{R}_s)$ satisfies $\dim \Null(Q(\mathscr{R}_s)) = 1$. 
\end{proposition}
\begin{proof}
Proposition \ref{prop:Qs_rank} established that 
\begin{align}\label{Si_def}
    p_i = S_i p_j,\quad S_i = \tau(\gamma_{ji}) \in O(2),
\end{align}
for all $i=1,\dots,n$, with $S_\l = I_2$.
The additional condition $(I_2-\tau(\gamma_\l))p_\l = 0$ implies that $p_\l \in  \mathcal{L}_\l$. 
There exists a unit vector $\hat{\mathcal{L}}_\l$ such that 
$p_\l = \alpha \hat{\mathcal{L}}_\l$ for some $\alpha\in\R$. Hence, $\Null(\mathcal{Q}(\mathscr{R}_s))=\IM(V_0)$ 
where 
\begin{align}\label{v0_def}
    V_0 =
\begin{bmatrix}
(S_1\hat{\mathcal{L}}_\l)^T &
\cdots &
(S_n \hat{\mathcal{L}}_\l)^T
\end{bmatrix}^T.
\end{align}

Therefore, $\dim\Null(\mathcal{Q}(\mathscr{R}_s))=1$.
\end{proof}

We now show that the control law drives the agents from any initial condition to a desired $\C_{nv}$-symmetric configuration.

\begin{theorem}\label{th_aug}
    Consider a MAS consisting of $n$ integrator agents \eqref{int-dynamics}, whose interaction topology is defined  by a spanning tree graph $\G_I\subset C_n$, and let
    $$\mathcal F = \{p \in \mathbb{R}^{2n} \, |\, \tau(\gamma)p_i=p_{\gamma(i)} \text{ for all } \gamma \in \C_{nv} \text{ and } i \in \V\}.$$
    Then, for any initial condition $p(0) \in \mathbb{R}^{2n}$,
    the control \eqref{ctrl_aug} renders the set $\mathcal{F}$ exponentially stable with $p(\infty)$ as the orthogonal projection of $p(0)$ onto $\mathcal{F}$,
\begin{align}\label{ts1}
    \lim_{t\to\infty}p(t)=\frac{1}{n}V_0V_0^{T}p(0),
\end{align}
where
\begin{align*}
    V_0=\begin{bmatrix}(S_1\hat{\mathcal{L}}_\l)^T &\cdots&(S_n \hat{\mathcal{L}}_\l)^T\end{bmatrix}^T\in\R^{2n},
\end{align*}
$\hat{\mathcal{L}}_\l$ is a unit direction vector along the mirror line $\mathcal{L}_\l$, and $S_i$ are defined in \eqref{Si_def}.
Furthermore, the steady-state of each agent is defined as 
\begin{align}\label{ts1_ea}
\lim_{t\rightarrow \infty}p_i(t)
    = \frac{1}{n} S_i\hat{\mathcal{L}}_\l \hat{\mathcal{L}}_\l^T
      \sum_{k=1}^n S_k^T p_k(0).
\end{align}
\end{theorem}
\begin{proof}
From Proposition~\ref{prop:Qs_aug_rank} we have
$$
    \Null\bigl(\mathcal{Q}(\mathscr{R}_s)\bigr)
    = \IM(V_0),
$$
where $V_0$ is given in \eqref{v0_def}. 
Then we have 
$$
V_0^T V_0
=
\sum_{i=1}^{n}
| S_i \hat{\mathcal{L}}_\l |^2=\sum_i1=n.
$$ 
Since $\mathcal{Q}(\mathscr{R}_s)$ is PSD and the columns of $V_0$ are orthogonal, we can define $\hat{V}$ as an orthonormal eigenbasis $\hat{V}=\begin{bmatrix}
    \hat{V}_0 & \hat{V}_+
\end{bmatrix}$ with $\hat{V}_0=\frac{1}
{\sqrt{n}}V_0$ and $V_+$ as the orthogonal complement of $V_0$. 
The proof continues in the same way as in Theorem \ref{th_s}, concluding with
\begin{align*}
    {\ \lim_{t\to\infty}p(t)=\hat{V}_0\hat{V}_0^Tp(0)=\frac{1}{n}V_0V_0^{T}p(0)\ },
\end{align*}
which shows \eqref{ts1}.
Also note that
$$
V_0^T p(0)=\sum_{k=1}^n (S_k\hat{\mathcal{L}}_\l)^T p_k(0)
=\sum_{k=1}^n \hat{\mathcal{L}}_\l^T S_k^T p_k(0).
$$
From  \eqref{ts1}, the $i$-th agent's component is therefore
$$
\lim_{t\to\infty}p_i(t)=\frac{1}{n} S_i\hat{\mathcal{L}}_\l\hat{\mathcal{L}}_\l^T \sum_{k=1}^n S_k^T p_k(0),
$$
which shows \eqref{ts1_ea}.

We now show that $p(\infty)$ lies in the set $\mathcal{F}$. From  \eqref{ts1} we have
$$
\alpha = \frac{1}{n} V_0^T p(0) \in \R,
$$
such that
\begin{align}\label{st_state}
    p(\infty) = \alpha V_0\in\IM(V_0), \quad p_i(\infty) = \alpha S_i \hat{\mathcal{L}}_\l,
\quad i\in\V.
\end{align}
 We are required to verify that $$
(I_2-\tau_\l)p_\l=0
\quad\Longleftrightarrow\quad
(I_2-\tau_i)p_i=0,
$$
for all $i \in \V$.
By definition, each agent’s mirror line satisfies
$$
\mathcal{L}_i = S_i \mathcal{L}_\l.
$$
Therefore, since $p_i=S_ip_\l$ and $p_\l\in\mathcal{L}_\l$. It follows that
$$
p_i \in S_i\mathcal{L}_\l=\mathcal{L}_i
\quad\Longleftrightarrow\quad
(I_2-\tau_i)p_i=0.
$$
So we have
$$E(\mathscr{R}_s)^T p(\infty) = 0,
\quad
(I_2-\tau(\gamma_i))\,p_i(\infty)=0.$$
Consequently, $p(\infty)$ satisfies all enforced reflectional relations and self-symmetry constraints. 
$$
p(\infty)\in\Null\big(Q(\mathscr{R}_s)\big).
$$
Additionally, we show that the steady-state
configuration $p(\infty)$ is invariant under every action in $\C_n$. This is expressed by the condition
$$
\tau(\rho)\,p_i(\infty)=p_{\rho(i)}(\infty),
\quad\forall\,i\in\V,\;\rho\in \C_n.
$$
To verify this identity, we use the group property
$$
\gamma_{\rho(i)\l}
=\rho\gamma_{i\l},
$$
yielding
$$
p_{\rho(i)}(\infty)=\tau(\gamma_{\rho(i)\l})p_\l
=\tau(\rho)\,\tau(\gamma_{i\l})p_\l
=\tau(\rho)\,p_i(\infty),
$$
which satisfies the rotational symmetry condition of a $\C_n$-symmetric configuration. This equivalently means
$$
E(\C_n)^T p(\infty) = 0,
$$
hence,
$$
p(\infty)\in\Null\big(Q(\C_n)\big).
$$

Finally, by definition, $\mathcal{F}=\Null\big(Q(\C_n)\big)\cap\Null\big(Q(\mathscr{R}_s)\big)$.
The relations established above for all
$\gamma\in \C_n$ and for all $\gamma\in\mathscr{R}_s$ imply that
$$
\tau(\gamma)p_i(\infty)=p_{\gamma(i)}(\infty)
\quad\text{for all }\gamma\in \C_{nv},\; i\in\V.
$$
Equivalently,  $p(\infty)\in F$. Therefore, Since $p(t)$ converges exponentially to $p(\infty)$, it follows that the set $F$ is exponentially stable, as claimed.
\end{proof}

\begin{example}\label{ex:c6_ex2}
Consider the same setup as in Example \ref{ex:c6_ex_r}, now are tasked with attaining a $\C_{nv}$-symmetric configuration solely under reflection constraints. Fig. \ref{fig:sim_c6_aug} shows the resulting trajectories and final configuration as a $\tau(\C_{nv})$-symmetric framework obtained with the augmented control law~\eqref{ctrl_aug}. 

\begin{figure}[h]
\begin{center}
\includegraphics[width=0.85\linewidth]{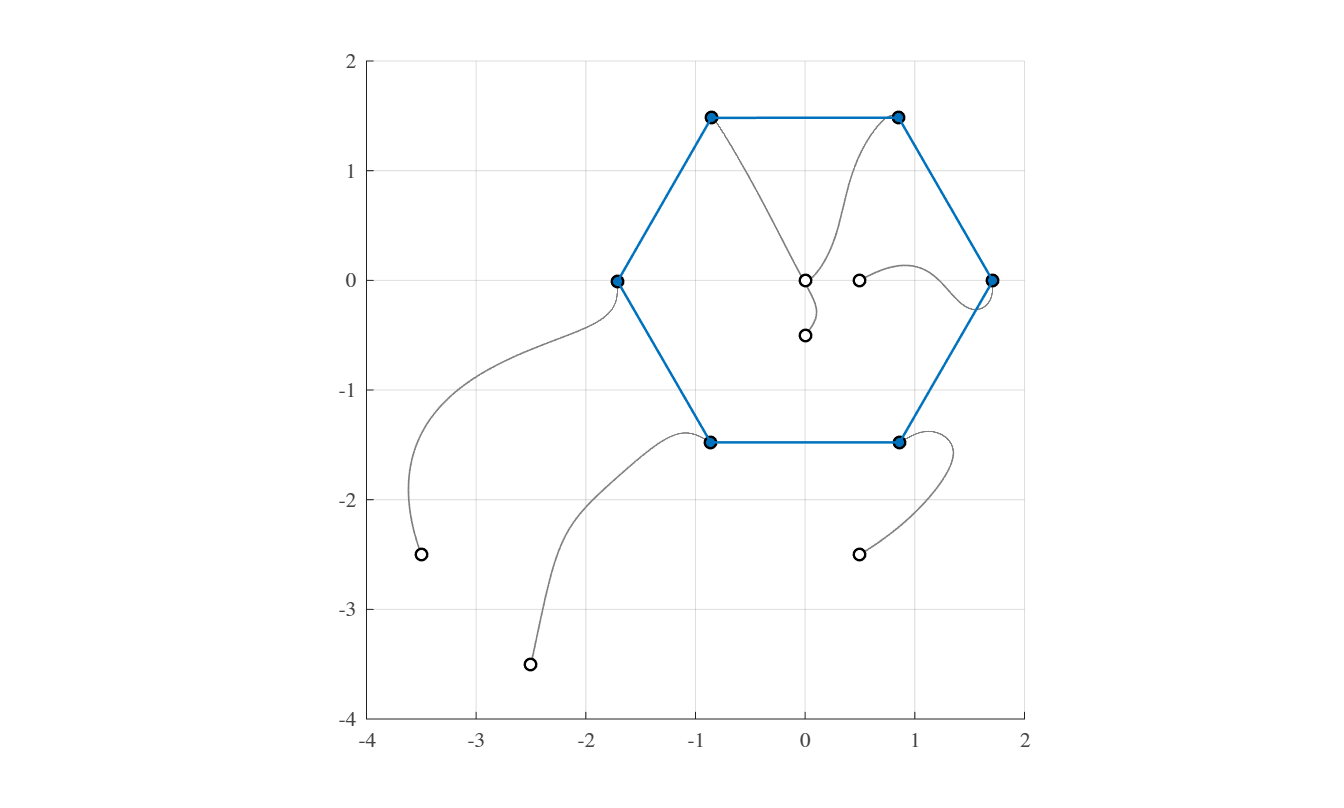}
    \end{center}
\vspace{-0.2cm}
\caption{Simulation results of the trajectories and final configuration of a $C_{6v}$-symmetric configuration under the augmented reflection-based control law \eqref{ctrl_aug}.}
\label{fig:sim_c6_aug}
\end{figure}
\end{example}

\section{Formation Maneuvering}\label{sec.maneuver}

Having achieved the reflection-based formation acquisition objective as shown in Fig. \ref{fig:sim_c6_aug}, note that the control law \eqref{ctrl_aug} successfully drives the agents to a desired configuration, but always with respect to a fixed inertial (global) origin and pre-defined orientation. In many practical scenarios, however, this may be limiting since the formation may require the ability to maneuver, that is, to translate, rotate, and scale while preserving the desired formation. 

To improve the flexibility of a symmetry-based formation control approach, 
(\cite{Zelazo2025forced}) proposed augmenting the closed-loop dynamics for each agent \eqref{ea_dyn} with a virtual state $r(t)$, enabling the agents to reach agreement on a different reference point. Building on this idea, (\cite{Martinez2025}) incorporated a virtual-state in a centralized formation maneuvering scheme, where all agents have access to a pre-defined trajectory, enabling the agents to converge to the desired configuration while maneuvering along the predefined trajectory. 

We now aim to extend this idea for the augmented reflection-based formation control strategy \eqref{ctrl_aug} as well.  In this direction, we define a virtual trajectory using the time-varying state $
        \chi(t) = \begin{bmatrix}
            r(t)^T & \theta(t)& s(t)
        \end{bmatrix}^T
    $
    where,  $r(t)\in\R^2$ is a translation, $\theta(t)\in \R$ is a rotation, and $s(t)\in\R^+$ is a reference scale factor.    The trajectory evolves according to
    $$
        \dot{r}(t)=v(t),\quad 
        \dot{\theta}(t)=\omega(t),\quad
        \dot{s}(t)=\alpha(t)s(t),
    $$
    where $v(t)\in\R^2$ is the translational velocity, $\omega(t)\in\R$ the angular velocity, and $\alpha(t)\in\R$ the  scaling rate of the desired formation.
    Let $\rr(\theta(t))\in SO(2)$ denote the associated rotation matrix. Its evolution satisfies, 
    $$
    \dot \rr(t)=\Omega(t)\rr(t),\quad\Omega(t)=\begin{bmatrix}0&-1\\1&0\end{bmatrix}\omega(t).
    $$
%
Since the symmetric formation achieved by the control law \eqref{ctrl_aug} is defined with respect to a fixed inertial point, we place the centroid of the desired formation at the origin. This serves as a natural reference for applying rotations and scalings to the formation. We then propose the augmented control law
\begin{align}\label{ctrl_mv}
    u(t)=&-Q(\ar_s, \theta(t))c(t)+\mathds{1}_n\!\otimes \! v(t) \nonumber\\&+(I_n \! \otimes \! \Omega(t)+\alpha(t)) c(t),
\end{align}
where $c(t)=p(t)-r(t)$, and $Q(\ar_s, \theta(t))$ has the block entries
{\small{$$ [Q(\ar_s, \theta(t))]_{ij} = \begin{cases}
                d(i)I_2, & i=j, \, i \in \V \\
                -\rr(\theta(t))\tau(\gamma_{ji})\rr(\theta(t))^T, & ij\in \mathcal{E}_I \\
                0, & \text{o.w.}
            \end{cases}.$$}}%
Here the reflection actions $\tau(\gamma_{ji})\in\ar_s$ undergo a \textit{similar transformation} ensuring that the reflection relations rotate consistently with the formation.
\begin{theorem}
 Consider a MAS consisting of $n$ integrator agents \eqref{int-dynamics}, whose interaction topology is defined  by a spanning tree graph $\G_I\subset C_n$, and let
    $$\mathcal F_c = \{p \in \mathbb{R}^{2n} \, |\, \tau(\gamma)c_i=c_{\gamma(i)} \text{ for all } \gamma \in \C_{nv} \text{ and } i \in \V\},$$
    be the set of all shifted $\mathcal{C}_{nv}$-symmetric configurations, with $c(t)=p(t)-r(t)$.
    Then, for any initial condition $p(0) \in \mathbb{R}^{2n}$,
    the control \eqref{ctrl_mv} renders the set $\mathcal{F}_c$ exponentially stable. 
\end{theorem}
\begin{proof}
Define $\zeta(t)\in\R^{2n}$ to be the configuration $p(t)\in\R^{2n}$ expressed in a frame moving along the virtual trajectory,
\begin{align}\label{moving_dynamics}
\zeta(t)=\frac{1}{s(t)}\big(I_n \otimes \mathcal{R}(t)^T\big)c(t).
\end{align}
We examine the derivative of each agent  $\zeta_i(t)\in\R^2$ (product rule),
{\small\begin{align*}
\dot{\zeta}_i(t)&=-\frac{\dot{s}(t)}{s^2(t)}\mathcal{R}(t)^Tc_i(t)+\frac{1}{s(t)}\big(\dot{\mathcal{R}}(t)^Tc_i(t)
     +\mathcal{R}(t)^T\dot{c_i}(t)\big).
\end{align*}}
Note that 
$\dot{\mathcal{R}}(t)^T=-\mathcal{R}(t)^T\Omega(t)$ and $\dot{\mathcal{R}}(t)^T=-\Omega(t)\mathcal{R}(t)^T$. Therefore 
\begin{align*}
    \dot{\zeta}_i(t)&=-\alpha(t)\zeta_i(t)-\Omega(t)\zeta_i(t)+\frac{1}{s(t)}\big(\mathcal{R}(t)^T(\dot{u}_i(t)-v(t))\big).
\end{align*}
By applying the control \eqref{ctrl_mv}, the dynamics for an agent $i$ are
\begin{align*}
    \dot{\zeta}_i(t)&=-\alpha(t)\zeta_i(t)-\Omega(t)\zeta_i(t)-\frac{1}{s(t)}\mathcal{R}(t)^Tv(t)
    \\
    &+\frac{1}{s(t)}\mathcal{R}^T\big(\sum_{{ij\in\E_I}}(\rr\tau({\gamma_{ji}})\rr^Tc_j(t)-c_i(t))
    \\
    &+v(t)+\Omega(t) c_i(t) + \alpha(t)c_i(t)\big).
\end{align*}
Since $\zeta_i(t)=\frac{1}{s(t)}\mathcal{R}(t)c_i(t)$, and $\tau({\gamma_{ji}})\rr^T=\rr\tau({\gamma_{ij}})$ all the trajectory dependent terms cancel, simplifying the expression to 
\begin{align*}
    \dot{\zeta}_i(t)=\sum_{{ij\in\E_I}}(\tau({\gamma_{ji}})\zeta_j(t)-\zeta_i(t)).
\end{align*}
This reduces to the analysis of the agents $\dot{\zeta}(t)=-Q(\ar_s)\zeta(t)$ to a static frame. By Theorem \ref{th_aug}, the dynamics of $\zeta(t)$ ensure that the formation exponentially converges to the set 
$$\mathcal{F}_\zeta= \{p \in \mathbb{R}^{2n} \, |\, \tau(\gamma)\zeta_i=\zeta_{\gamma(i)} \text{ for all } \gamma \in \C_{nv} \text{ and } i \in \V\},$$
From the definition of $\zeta_i(t)$ \eqref{moving_dynamics}, this set is equivalent to $\mathcal{F}_c$, rendering the set $\mathcal{F}_c$ exponentially stable, as claimed.
\end{proof}

Therefore, since $c(t)=p(t)-r(t)$ at any given time, and the control law \eqref{ctrl_mv} ensures that $c(t)$ converges to a $\C_{nv}$-symmetric configuration, the agents $p(t)$ themselves converge exponentially to the desired symmetric formation shifted by $r(t)$.

\begin{example}\label{ex:c6_ex_mv}
Consider a MAS consisting of $n=4$ agents, tasked with attaining a $\C_{4v}$-symmetric configuration (see Fig. \ref{fig:sym_fw_Cn}(b)) solely under reflection constraints. The agents are simultaneously tasked to move as a unit following a pre-defined trajectory to avoid an obstacle and maintain a specific orientation about it.
Fig.~\ref{fig:sim_c4_mv} illustrates the resulting trajectories and the final configuration obtained under the augmented control law~\eqref{ctrl_mv}.

\begin{figure}[h]
\begin{center}
\includegraphics[width=0.98\linewidth]{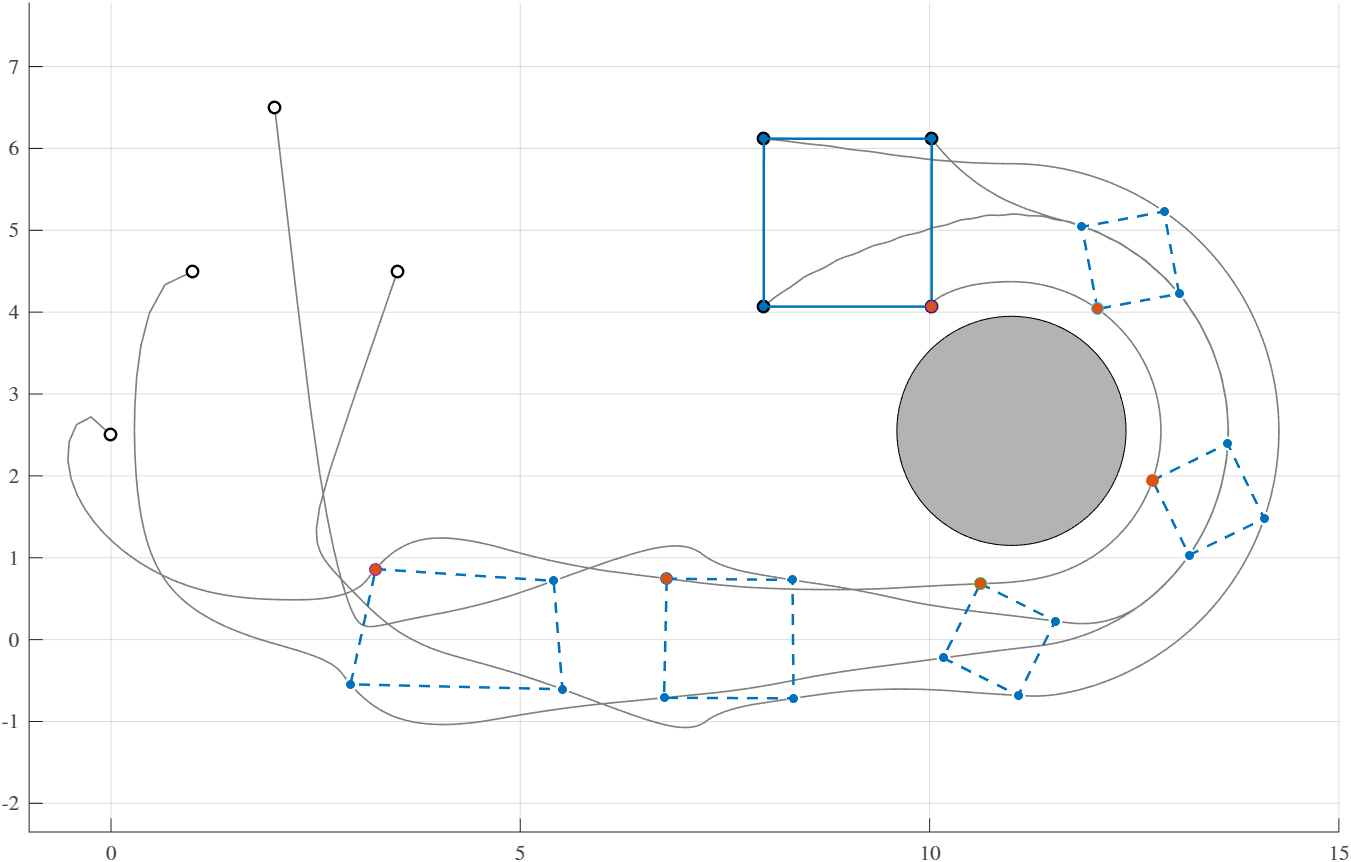}
    \end{center}
\vspace{-0.2cm}
\caption{Simulation results of the trajectories and time-varying configuration of a $\C_{4v}$-symmetric formation generated from \eqref{ctrl_mv}.}
\label{fig:sim_c4_mv}
\end{figure}
\end{example}

\section{Conclusion}

This paper introduced a symmetry-based formation control method in $\R^2$ that relies solely on inter-agent reflection relations together with a single anchored agent. We showed that any $\C_{nv}$-symmetric formation can be achieved using only $n-1$ communication links, with guaranteed stability and convergence. The approach was further extended to a reflection-constrained control law preserves the target configuration while maneuvering along a pre-defined trajectory. These results highlight a potential alternative to rigidity-based formation control, and future directions include extending the framework to point group symmetries in $\R^3$, analyzing more general information structures such as directed and switching graphs, and incorporating leader–follower structures to generalize the approach and enable fully distributed maneuvering.

{\small \bibliography{references}}

\end{document}